\documentclass[12pt]{article}
\usepackage{mathrsfs}
\usepackage{theorem}
\usepackage[latin1]{inputenc}
\usepackage[center]{caption}
\usepackage{amsfonts}
\usepackage{amstext}
\usepackage{amsmath}
\usepackage{amssymb}
\usepackage{graphicx}
\usepackage{epic}
\usepackage{psfrag}
\usepackage{scrpage2}

\newcommand{\Vt}{\widetilde V}
\newcommand{\beq}{\begin{equation}}
\newcommand{\eeq}{\end{equation}}
\newcommand{\be}{\begin{equation}}
\newcommand{\ee}{\end{equation}}
\newcommand{\ba}{\begin{array}}
\newcommand{\ea}{\end{array}}
\newcommand{\beqa}{\begin{eqnarray}}
\newcommand{\eeqa}{\end{eqnarray}}
\newcommand{\bea}{\begin{eqnarray}}
\newcommand{\eea}{\end{eqnarray}}

\newcommand{\cO}{{\cal O}}
\newcommand{\cB}{{\cal B}}

\newcommand{\cL}{{\cal L}}
\newcommand{\re}{{\rm Re}}

\newcommand{\no}{\nonumber}

\newcommand{\YU}{Y_u}

\newcommand{\Vtr}{\widetilde V_0}
\newcommand{\ts}{\tilde s}
\newcommand{\tc}{\tilde c}

\newcommand\pubnumber{TUM-HEP-778/10}
\newcommand\pubdate{\today}

\def\munch{Physik Departnment, Technische Universit\"at M\"unchen\\ D-85748 Garching, Germany}
\def\mail{\footnote{katrin.gemmler@ph.tum.de}}

\def\Title#1{\begin{center} {\Large #1 } \end{center}}
\def\Author#1{\begin{center}{ \sc #1} \end{center}}
\def\Address#1{\begin{center}{ \it #1} \end{center}}

\newcommand\pubblock{\rightline{\begin{tabular}{l} \pubnumber\\
         \pubdate  \end{tabular}}}
\newenvironment{Abstract}{\begin{quotation}  }{\end{quotation}}
\newenvironment{Presented}{\begin{quotation} \begin{center} 
             PRESENTED AT\end{center}\bigskip 
      \begin{center}\begin{large}}{\end{large}\end{center} \end{quotation}}
\def\Acknowledgements{\bigskip  \bigskip \begin{center} \begin{large}
             \bf ACKNOWLEDGEMENTS \end{large}\end{center}}

\begin{document}
\begin{titlepage}
\pubblock

\vfill
\Title{An effective look at\\ right-handed currents and quark flavour mixing}
\vfill
\Author{ Katrin Gemmler\mail}
\Address{\munch}
\vfill
\begin{Abstract}
We present an effective theory approach for a model with a left-right symmetric flavour group 
broken only by the Yukawa couplings. An underlying $SU(2)_L \times SU(2)_R \times U(1)_{B-L}$ global symmetry is assumed without specifying the fundamental theory. The model is motivated by the fact that the $|V_{ub}|$ problem can be solved by assuming the existence of right-handed (RH) currents.  In this framework the flavour mixing in the RH sector is given by a new mixing matrix. After a short introduction of the model, we show how this RH mixing matrix can be determined by theory, experimental data and phenomenological constraints. We summarize the results of the RH contribution to flavour mixing in $\Delta F=2$ processes. Furthermore we briefly discuss $Z\to b \bar b$ and the rare decays $B_{s,d}\to \mu^+\mu^-$, $B\to \{X_s,K,K^*\} \nu\bar\nu$ and $K\to \pi\nu\bar\nu$ with corresponding correlations.
\end{Abstract}
\vfill
\begin{Presented}
Proceedings of CKM2010, the 6th International Workshop on the CKM Unitarity Triangle,\\ University of Warwick, UK, 6-10 September 2010
\end{Presented}
\vfill
\end{titlepage}
\def\thefootnote{\fnsymbol{footnote}}
\setcounter{footnote}{0}

\section{Introduction}

Right-handed currents are known to arise in various new physics scenarios, 
particularly in models with an underlying $SU(2)_R \times SU(2)_L$ symmetry. 
%[?]
However here we follow a different approach.
Recently, right-handed currents have also been shown to resolve the tension 
between inclusive and exclusive determinations of the  $|V_{ub}|$ element of 
the CKM matrix (for example see \cite{Crivellin:2009sd}). Motivated by this work we consider RH currents 
in an effective theory approach, assuming a left-right symmetric flavour group which is broken only by the Yukawa couplings. A new unitary matrix controlling flavour-mixing 
in the RH sector appears, whose structure can be determined by 
charged-current data. We analyse the impact on neutral-current 
flavour-violating processes in particle-antiparticle mixing, $Z\to b \bar b$ 
and various rare decays. A more detailed analysis can be found in \cite{Buras:2010pz}.

\section{The RHMFV model}

\begin{figure}[h]
\begin{center}
\includegraphics[width=7cm]{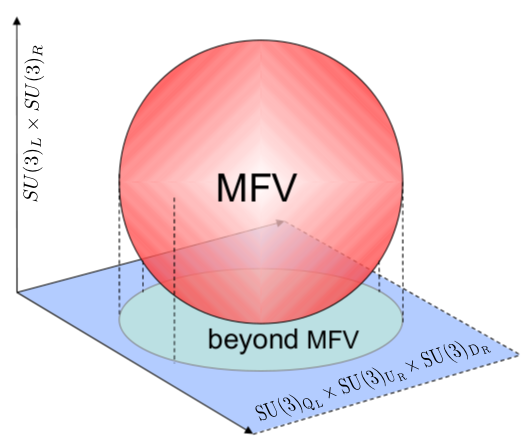}
\end{center}
\caption{}
\label{fig:com} 
\end{figure}

In the Right-handed Minimal Flavour Violation (RHMFV) model introduced by \cite{Buras:2010pz} an effective theory approach is considered in the spirit of \cite{D'Ambrosio:2002ex}. However, we choose a left-right (LR) symmetric flavour symmetry $SU(3)_L \times SU(3)_R $ which is broken just by the Yukawas, similar to the MFV case. In this approach the fundamental theory is not specified, and we only make assumptions about the global symmetry and the pattern of breakdown. In order to make the minimal model, the Standard Model (SM) gauge group is embedded into a global $SU(2)_L\times SU(2)_R \times U(1)_{B-L}$ group, where only $SU(2)_L$ and $U(1)_Y$ are effectively gauged below the TeV scale.
Figure \ref{fig:com} compares our model with respect to MFV. RHMFV is clearly beyond MFV if considered in comparison to the $SU(3)^3$ quark flavour symmetry of MFV. However if we go to the LR symmetric flavour group, then it is MFV in a sense as the symmetry is again just broken by the Yukawas. 

In the next step all dimension six operators formally invariant under the LR symmetric flavour group are considered. New bilinears arise with respect to MFV e.g. $\bar Q_R  \Gamma  \YU^\dagger\YU Q_R$. In this bilinear the Yukawa insertion $\YU^\dagger\YU$ is present, characterizing the strength of RH mediated flavour changing neutral currents
(FCNCs) and containing elements of a new RH mixing matrix $\Vt$. 
This matrix $\Vt$ appears due to misalignment of the Yukawas in the down-type sector and controls the RH flavour mixing. 
It can be parametrized as follows
\begin{equation}
\Vt = D_U \Vtr D^\dagger_D\,,
\label{param}
\end{equation}
where $\Vtr$ is a CKM like matrix and $D_{U,D}$ are two diagonal matrices, with five more new CP violating phases in particular $D_U={\rm diag}(1, e^{i\phi^u_2}, e^{i\phi^u_3})$ and 
$D_D={\rm diag}(e^{i\phi^d_1}, e^{i\phi^d_2}, e^{i\phi^d_3})$. Altogether there are three new real mixing angles and six new complex phases. 
This matrix is restricted by various bounds, 
%in particular from charged current data, unitarity and phenomenology like FCNCs, 
which I want to discuss in the following in detail . 
 
Using the data on tree level charged current transitions, in particular $u\to d$,
 $u\to s$, $b\to u$ and $b\to c$, allows us to set bounds on four elements of $\Vtr$. 
\be
| \Vt | \sim  
\left(\begin{array}{ccc}
 < 1.4~ &  < 1.4~ &  1.0\pm 0.4 ~ \\ 
- & - &    < 2.0 ~ \\ - & - &  - 
\end{array}\right) \times \left(\frac{10^{-3}}{\epsilon_R}\right)\,.
\ee
Note that elements of $\Vt$ and $\epsilon_R$ appear only in combination, where $\epsilon_R$ gives the size of the effective RH charged current coupling (normalized to the SM left-handed one).
Unitarity of the first row puts a significant constraint on the value of $\epsilon_R$ 
\be
|\epsilon_R|  =   \left( |\epsilon_R \Vt_{ud}|^2 + |\epsilon_R \Vt_{us}|^2 
+ |\epsilon_R \Vt_{ub}|^2  \right)^{1/2} = (1.0 \pm 0.5) \times 10^{-3}\, ,
\ee
being in agreement with the result from naive estimates using coupling coefficients of $\cO(1)$ and a scale of $\Lambda \approx 3$~TeV. From the unitarity of the third column one can deduce that the large $|\Vt_{ub}|$ constrains the maximal value of $|\Vt_{tb}|$. A large value of $|\Vt_{tb}|$ is welcome since it minimizes the contribution to the elements $|\Vt_{ts}|$ and $|\Vt_{td}|$ contributing to FCNCs which are highly constrained. Also it could help to improve the agreement with $Z\to b \bar b$ in the RH sector with experiment. A global fit maximizing $|\Vt_{tb}|$ is performed. From this fit we conclude that the RH mixing matrix is well described by the following ansatz
\be
 \Vtr^{\rm (II)} = 
\left(\begin{array}{ccc}
\pm  \tc_{12} \frac{\sqrt{2}}{2} & \pm \ts_{12} \frac{\sqrt{2}}{2} & - \frac{\sqrt{2}}{2} \\ 
-\ts_{12} &\tc_{12} &  0 \\ 
\tc_{12} \frac{\sqrt{2}}{2} & \ts_{12} \frac{\sqrt{2}}{2} &  \pm \frac{\sqrt{2}}{2}
\end{array}\right)\,,
\ee
where we have only one free parameter.

\begin{figure}[t]
\begin{center}
\includegraphics[width=8cm]{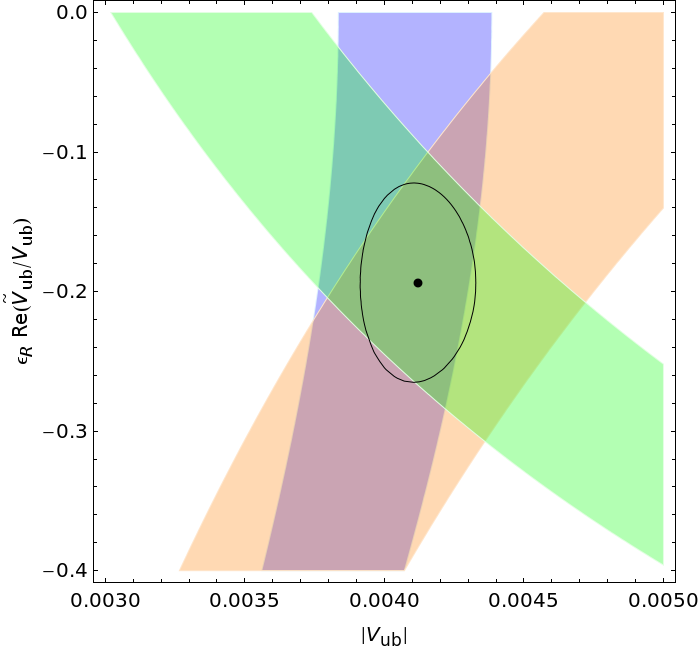}
\vskip -7.4 cm 
{\footnotesize $\qquad \qquad  B\to \pi \ell \nu \qquad
B\to X_u \ell \nu \qquad B\to \tau\nu$}
\vskip  6.9 cm 
\end{center}
\caption{\label{fig:Vub} Constraints on $|V_{ub}|$ and 
$\epsilon_R~ \re\left(\frac{ \Vt_{ub} }{ V_{ub}}\right)$
from $B\to \pi \ell \nu$ (green), $B\to X_u \ell \nu$ (blue), and
$B\to \tau\nu$ (orange). 
The bands denote the $\pm 1 \sigma$ intervals of the various experimental
constraints. The ellipse denotes the $1 \sigma$ region of our 
best-fit solution.}
\end{figure}

Let us have a closer look at the $\Vt_{ub}$ element. In order to set a bound on $\epsilon_R \Vt_{ub}$ we compare the effectively appearing CKM combinations to the values derived from the comparison of the SM with experiment. 
The structure of combinations of the CKM matrix and the new RH mixing matrix is easy to derive as the decay $B \to \pi \ell \nu$ is only sensitive to vector current, while $B\to \tau\nu$ is only sensitive to axial current. In the inclusive decay, a mixing between left and right components appears, which however turn out to be
negligible. This allows us to set up three conditions. The global fit solution shown in figure \ref{fig:Vub}, shows that in the presence of RH currents the discrepancy between inclusive and exclusive determinations is resolved. This is not the case in the SM case which corresponds to the top of the vertical axis, where the 
three determinations of $|V_{ub}|$ are clearly different.
In presence of RH currents the true value of $|V_{ub}|$ turns out to be $(4.1 \pm 0.2 )\times 10^{-3}$, selecting the inclusive determination as the true value.

\section{\boldmath$\Delta F=2$ Processes}
After having fixed the RH mixing matrix, we want to examine whether this new kind of mixing is consistent with low energy observables from particle anti-particle mixing and rare decays.

Let us first have a closer look at $\Delta F=2$ processes.
Within our framework new operators are generated, where we focus on contributions of the $\bar Q_R \YU^\dagger \YU \gamma^\mu Q_R$ bilinear, in particular
\bea
\mathcal{O}^{(6)}_{RR} &=&  [\bar Q^i_R (\YU^\dagger \YU)_{ij} \gamma_\mu Q^j_R]^2 \nonumber  \\
\mathcal{O}^{(6)}_{LR} &=&  [\bar Q^i_L (\YU \YU^\dagger)_{ij} \gamma^\mu  Q^j_L]
[\bar Q^i_R (\YU^\dagger \YU)_{ij} \gamma_\mu Q^j_R] \,.
\eea
This RR and especially the LR structure are known to be dangerous due to renormalization group effects (and in the case of $\varepsilon_K$ also through chiral enhancement). The effective Hamiltonian reads
\be
\cL^{\Delta F=2} = \frac{c_{RR}}{\Lambda^2} \mathcal{O}^{(6)}_{RR} +
\frac{c_{LR}}{\Lambda^2} \mathcal{O}^{(6)}_{LR} \,,
\label{lag}
\ee
where $c_{RR}$ and $c_{LR}$ are flavour-blind dimension-less coefficients, whose size will be discussed later. Hence the RH mixing is only determined by the elements of the RH mixing matrix in particular by $\tc_{12}$, $\ts_{12}$ and CP violating phases.

%The corresponding LL operator would be already present in the MFV framework contributing to
%short-distance corrections only helicity structure and CKM factors of the SM.
%As a result, it cannot modify the SM predictions for the time-dependent CP asymmetries in $B$ decays e.g. in $S_{\psi K}$ and  $S_{\psi\phi}$.

How does this RH mixing matrix now contribute to the different processes? The structure is summarized in table \ref{tab1}.
\begin{table}[htbp]
\renewcommand{\arraystretch}{1.4}
\centering
\begin{tabular}{|l|c|c|c|}
\hline 
{Mixing term} & $K$-mixing &  $B_d$-mixing & $B_s$-mixing \\
& $s\to d$  & $b\to d$ & $b\to s$ \\
\hline 
{  $\Vt^*_{ti} \Vt_{tj}$  } 
&  $ \frac{1}{2} \tc_{12} \ts_{12} e^{i(\phi^d_2-\phi^d_1)} $ 
&  $ \pm \frac{1}{2} \tc_{12} e^{i(\phi^d_3-\phi^d_1)} $ 
&  $ \pm \frac{1}{2} \ts_{12} e^{i(\phi^d_3-\phi^d_2)} $ \\
\hline
\end{tabular}
\caption{}
\label{tab1}
\renewcommand{\arraystretch}{1.0}
\end{table}
The strong constraints of the $K$ system mainly by $\varepsilon_K$ point towards either a small $\tc_{12}$ or $\ts_{12}$ unless  $c_{RR}$ and $c_{LR}$ are very small.
Due to the enhanced value of $S_{\psi\phi}$ from CDF \cite{Aaltonen:2007he} and D0 \cite{:2008fj} collaborations, we assume large CP-violating effects in $B_s$-mixing. Hence we choose $\ts_{12}$ to be large, which then in combination with $\varepsilon_K$ implies automatically negligible effects in $B_d$ mixing. Roughly the RH mixing matrix will appear with the following structure
\be
\left| \Vtr \right|  \sim
\left(\begin{array}{ccc}
   0  & \frac{\sqrt{2}}{2} &  \frac{\sqrt{2}}{2} \\ 
   1  &  0  &  0 \\ 
   0  & \frac{\sqrt{2}}{2} &  \frac{\sqrt{2}}{2}
\end{array}\right)\,,
\ee
where the zero entries shouldn't be understood as exact zeros. Note that there are additional phases from the diagonal matrices shown in equation (\ref{param}).

For a more detailed analysis we can consider the constraints from $\varepsilon_K$ and $B_s$ mixing simultaneously. Imposing 
\be
 \frac{ (\Delta M_s)_{\rm exp} }{ (\Delta M_s)_{\rm SM} } 
\approx 0.96 \pm 0.15 \qquad  \text{and} \qquad
 S_{\psi \phi } 
\approx 0.6 \pm 0.3\,,
\ee 
these two conditions for $B_s$ mixing allow us to set up a system of two equations which can be solved. One obtains conditions for coupling coefficients and phases, which we will apply in the latter analysis of rare decays. For example assuming the RR operator in (\ref{lag}) to be dominant ($c_{RR} \gg c_{LR}$), the solutions are given by
\bea
&& c_{RR} \approx \pm 7.3 \times 10^{-3} \quad {\rm and} \quad \sin(2\phi^d_{32}) \approx \mp 0.30~, \no \\
&& c_{RR} \approx \pm 2.3 \times 10^{-3} \quad {\rm and} \quad \sin(2\phi^d_{32}) \approx \mp 0.95~.
\label{sol}
\eea
\begin{table}[htbp]
\centering
\begin{tabular}{|l|l|l|}
\hline 
{operator} &  {size of coefficient}  &  {suppression} \\
\hline 
RH charged current   & $\cO(1)$ &tree level \\
$\Delta F=2$ & $1/(16 \pi^2)\approx 6 \times 10^{-3}$ &loop  \\
\hline
\end{tabular} 
\caption{}
\label{tab2}
\end{table}
We obtain coupling coefficients of $\cO(10^{-3})$ which are substantially lower than the $\cO(1)$ Wilson coefficients determined from charged-currents. 
Naively the small value of $c_{RR}$ doesn't seem natural, however this is perfectly fine when we understand $\Delta F=2$ operators to be loop-suppressed with respect to charged current ones. This situation is summarized in table \ref{tab2}. The solutions for the parameters from $B_s$ mixing also satisfy the kaon bounds.
%In case of the LR operator providing the dominant contribution ($c_{RR} \ll c_{LR}$) to $B_s$ mixing more fine-tuning is required due to large chiral enhancement of the left-right operator in $\varepsilon_K$.

\section{Effects due to \boldmath$\sin(2\beta)$ enhancement}
Let us have a closer look at $\varepsilon_K$. As RH currents choose the inclusive determination giving the ``true`` $|V_{ub}|$, the value for $\sin(2\beta)^{\rm RH}_{\rm tree} = 0.77 \pm 0.05$ obtained from tree-level determinations is substantially higher than the corresponding result obtained in the SM, where the inclusive and exclusive determinations of $|V_{ub}|$ are averaged. It follows that the tension between the experimental value of $\varepsilon_K$ and its prediction within the SM is automatically resolved within our framework.
However as the NP contribution to $B_d$ mixing is constrained to be negligible, we expect $S_{\psi K_S}^{\text{RH}}$ to be significantly larger than the experimental value $S_{\psi K_S}^{\text{exp}}= 0.672 \pm 0.023$ \cite{Barberio:2008fa}.
As a result the existing tension between these two values increases. Our result is about $2\sigma$ larger than the measured value. That means that $S_{\psi K_S}$ cannot be solved by RH currents alone in this framework. 
However a significant lower measured value of $S_{\psi\phi}$ might allow effects in $B_d$ mixing, which could lead to a high enough new phase in $B_d$ mixing and then ease the tension.

\section{Analysis of rare decays and \boldmath$Z\to b\bar b$}
We now take a look at $Z\to b\bar b$ and rare decays. The main contribution is arising from a new effective coupling from the $Z$ boson to RH down type quarks. We have applied our analysis to a number of processes of which I will here only present a selection. First of all we find that the constraints from $B_{s,d} \to \ell^+\ell^-$ eliminate the possibility of removing the known anomaly $Z\to b\bar b$ in the RH sector. Furthermore the constraint from $B_s\to X_s \ell^+\ell^-$ precludes $B_{s}\to \mu^+\mu^-$ near present experimental bound. The deviations in $\cB(B_{s}\to \mu^+\mu^-)$ can still be $\cO(1)$, but push the effects in $\cB(B_{d}\to \mu^+\mu^-)$ to be small or negligible when using the constraints from  $S_{\psi\phi}$.

\begin{figure}[t]
\begin{center}
\includegraphics[width=7cm]{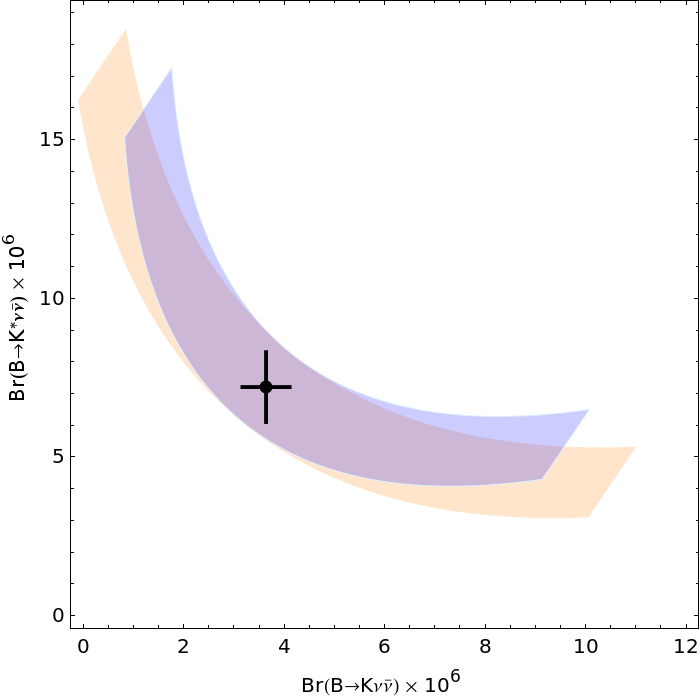}\qquad \quad
\includegraphics[width=7cm]{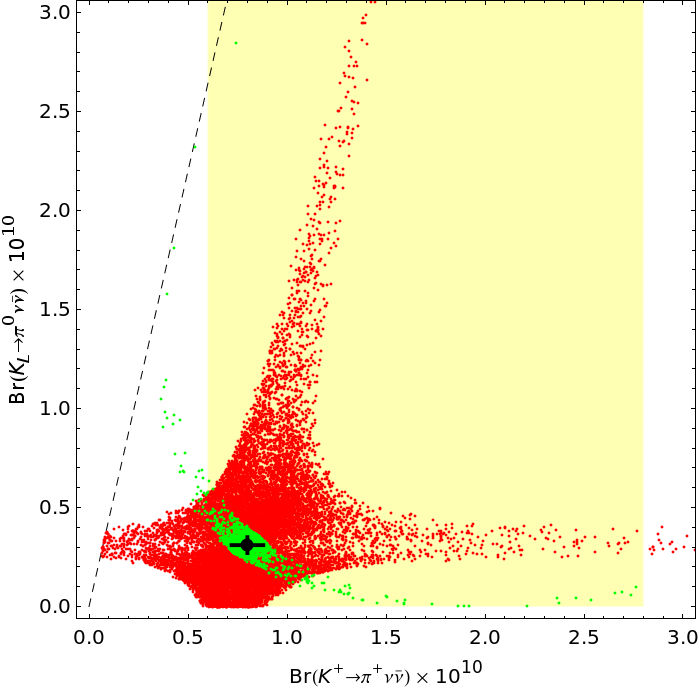}
\end{center}
\caption{
left: Correlation between
$\cB(B\to K \nu \bar \nu)$ and 
$\cB(B\to K^* \nu \bar \nu)$,\newline
right: Correlations between 
$\cB(K^+ \to \pi^+ \nu \bar \nu)$ and 
$\cB(K_L \to \pi^0 \nu \bar \nu)$
}
\label{fig3} 
\end{figure}

In addition interesting correlations can be found. Figure \ref{fig3} shows the correlation between $\cB(B\to K \nu \bar \nu)$ and $\cB(B\to K^* \nu \bar \nu)$. We show two bands
corresponding to the values of equation (\ref{sol}), a blue (dark gray) band for $|\sin(2\phi^d_{32})|=0.95$ and an orange (light gray) band for $|\sin(2\phi^d_{32})|=0.30$). We observe a clear anti-correlation. Both branching ratios can be enhanced by more than a factor of two over the SM value, which is shown as a black dot with the corresponding error bars.

Also a specific pattern is obtained for the correlations between $\cB(K^+ \to \pi^+ \nu \bar \nu)$ and $\cB(K_L \to \pi^0 \nu \bar \nu)$ imposing the constraints from $\varepsilon_K$ and $S_{\psi\phi}$. This correlation is shown in Figure \ref{fig3}. Depending on which operator dominates $S_{\psi\phi}$ we obtain a different correlation due to different conditions on the phases. 
Assuming $c_{LR}$ to be dominant (shown in green/light points) the situation is more constrained. For dominant $c_{RR}$ (shown in red/dark points) $\cO(1)$ deviations from the SM are possible. With larger deviations we end up in a more fine-tuned scenario. The phase of $\phi^d_{12}$ has to be tuned in order to satisfy the  $\varepsilon_K$ constraint. As noted by \cite{Blanke:2009pq}, this structure is characteristic for all NP frameworks where the phase in  $\Delta S=2$ amplitudes is the square of the CP-violating phase in $\Delta S=1$ FCNC amplitudes. 

\section{Conclusions}
Summarizing in short the advantages and disadvantages of the RHMFV model.
RH currents provide a solution to the $V_{ub}$ problem. In addition the $S_{\psi\phi}$ and $\varepsilon_K$ anomalies can be understood. However the $Z b\bar b$ problem cannot be resolved.
The tension between $\sin 2\beta$ and $S_{\psi K_S}^{\text{exp}}$ becomes stronger due to the fact that we implemented a large $S_{\psi\phi}$. Furthermore, when taking $S_{\psi\phi}$ large,  the effects in both $B_d$ mixing and rare $B_d$ decays turn out to be negligible. A well-defined pattern of correlations were obtained in various decays.

\Acknowledgements
I warmly thank Andrzej Buras and Gino Isidori for the very pleasant collaboration.
This research was supported by the German Bundesministerium f\"ur Bildung und Forschung under contract 05H09WOE.

\end{document}